\documentstyle[12pt,aasms4]{article}
\newcommand{\postscript}[2]
 {\setlength{\epsfxsize}{#2\hsize}
  \centerline{\epsfbox{#1}}}
\def\tempest%
{\begin{array}{ccc}
1 & 1 & 1 \\
1 & 1 & 1 \\
4 & 3 & 8
\end{array}}

\begin{document}

\title{The Intrinsic Shapes of Low-Surface-Brightness Dwarf Irregular\\
        Galaxies and Comparison to Other Types of Dwarf Galaxies}
\bigskip
\author{Eon-Chang Sung}
\affil{Korea Astronomy Observatory, Taejon, Korea 305-348, \\
       Department of Astronomy, Yonsei University, Seoul, Korea 120-749 \\
       ecsung@hanul.issa.re.kr,}
\authoremail{ecsungi@hanul.re.kr}
\bigskip

\author{Cheongho Han}
\affil{Department of Astronomy \& Space Science, \\
       Chungbuk National University, Cheongju, Korea 361-763 \\
       cheongho@astro-3.chungbuk.ac.kr,}
\authoremail{cheongho@astro.chungbuk.ac.kr}
\bigskip

\author{Barbara S. Ryden}
\affil{Department of Astronomy, The Ohio State University, Columbus, OH 43210 \\
       ryden@astronomy.ohio-state.edu,}
\authoremail{ryden@astronomy.ohio-state.edu}
\bigskip

\author{Richard J.\ Patterson}
\smallskip
\affil{Astronomy Department, University of Virginia, 
       Charlottesville, VA, 22903 \\
       ricky@virginia.edu,}
\authoremail{ricky@virginia.edu}
\bigskip

\author{Mun-Suk Chun}
\smallskip
\affil{Department of Astronomy, Yonsei University, Seoul, Korea 120-749 \\
       mschun@galaxy.yonsei.ac.kr,}
\authoremail{mschun@galaxy.yonsei.ac.kr}
\bigskip

\author{Woo-Baik Lee}
\affil{Korea Astronomy Observatory, Taejon, Korea 305-348 \\
       wblee@hanul.issa.re.kr,}
\authoremail{ecsungi@hanul.re.kr}
\bigskip

\author{Dong-Jin Kim}
\smallskip
\affil{Department of Astronomy \& Space Science, \\
       Chungbuk National University, Cheongju, Korea 361-763 \\
       cheongho@astro-3.chungbuk.ac.kr,}
\authoremail{kimdj@astro.chungbuk.ac.kr}
\bigskip

\begin{abstract}
  
  In this paper, we measure the ellipticities of 30 LSB dI galaxies and
  compare the ellipticity distribution with that of 80 dEs (Ryden \& Terndrup
  1994; Ryden et al.\ 1998)\markcite{ryden 1994; 1998} and 62 BCDs (Sung et
  al.\ 1998).\markcite{sung1998} We find that the ellipticity distribution of
  LSB dIs is very similar to that of BCDs, and marginally different from that
  of dEs.  We then determine the distribution of intrinsic shapes of dI
  galaxies and compare to those of other type dwarf galaxies under various
  assumptions.  First, we assume that LSB dIs are either all oblate or all
  prolate, and use non-parametric analysis to find the best-fitting
  distribution of intrinsic shapes. With this assumption, we find that the
  scarcity of nearly circular LSB dIs implies, at the 99\% confidence level,
  that they cannot be a population of randomly oriented oblate or prolate
  objects.  Next, we assume that dIs are triaxial, and use parametric analysis
  to find permissible distributions of intrinsic shapes.  We find that if the
  intrinsic axis ratios, $\beta$ and $\gamma$, are distributed according to a
  Gaussian with means $\beta_0$ and $\gamma_0$ and a common standard deviation
  of $\sigma$, the best-fitting set of parameters for LSB dIs is
  $(\beta_0,\gamma_0,\sigma) = (0.66,0.50,0.15)$, and the best fit for BCDs is
  $(\beta_0,\gamma_0,\sigma) = (0.66,0.55,0.16)$, while the best fit for dEs
  is $(\beta_0,\gamma_0,\sigma) = (0.78,0.69,0.24)$.  The dIs and BCDs thus
  have a very similar shape distribution, given this triaxial hypothesis,
  while the dEs peak at a somewhat more spherical shape. Our results are
  consistent with an evolutionary scenario in which the three types of dwarf
  galaxy have a close relation with each other.

\end{abstract}

\vskip20mm

\centerline{submitted to {\it The Astrophysical Journal}: ??? ??, 1998}
\centerline{Preprint: CNU-A\&SS-01/98}
\clearpage

\section{Introduction}

Although the faint end of the galaxy luminosity function is not well
determined, recent studies indicate that low-surface-brightness (LSB) dwarf
galaxies are by far the most numerous type of galaxy, and contribute a
significant fraction of the mass of the universe (Reaves 1983; Binggeli,
Sandage, \& Tammann 1985; Phillipps et al.\ 1987)\markcite{reaves1983,
sandage1985, phillipps1987}.  Morphologically, dwarf galaxies, like their
counterpart bright galaxies, are classified into several types.  The most
common type of dwarf galaxy ($\sim 80\%$ of the total) is the dwarf elliptical
(dE). These galaxies have regular elliptical isophotes and roughly exponential
surface brightness profiles; they are often found in groups and clusters
(Davies et al.\ 1988).\markcite{davies1988} The second type of dwarf galaxy is
the blue compact dwarf (BCD) galaxy.  In contrast to gas-poor dEs, BCDs
contain giant HII regions surrounding O and B stars within a massive HI
reservoir; BCDs exhibit spectra slowly rising toward the blue, implying that
they are undergoing intense star formation (du Puy 1970; Searle \& Sargent
1972).\markcite{dupuy1970, searle1972} Most BCDs have regular isophotes in the
outer region, like dEs, but the inner isophotes are frequently distorted from
ellipses, due to the presence of bright HII regions (Loose \& Thuan
1986).\markcite{loose1986}

The final type of dwarf galaxy is the LSB dwarf galaxy.
LSB dwarfs include both irregular (dI) and more regular
spiral (dS) galaxies.
Like BCDs, they contain a large amount of HI, often with small 
OB associations, and have blue colors ($B-V\sim 0.5\ {\rm mag}$),
indicating a significant level of recent star formation
(Staveley-Smith, Davies, \& Kinman 1992)\markcite{staveley1992}. 
However, they are distinguished from BCD galaxies
by their amorphous shapes even in the outer region.
Additionally, in contrast to dEs, these galaxies are more likely to be found
outside of clusters (Bingelli, Tarenghi, \& 
Sandage 1990).\markcite{bingelli1990}

The evolutionary connections among the three different types of dwarf galaxies
remain both elusive and confusing.  There are two major competing hypotheses
for the evolutionary connection between BCDs and dEs.  The first hypothesis
claims that BCDs are basically a different population from dEs, as evidenced
by the spectroscopic and spectrophotometric differences.  According to this
scenario, BCDs are truly young systems, in which the present star burst is the
first in the galaxy's lifetime.  The second hypothesis suggests that BCDs,
like dEs, are mainly composed of old stellar populations, and that their
observed spectroscopic features and spectral energy distributions are the
result of a recent burst of star formation 
(Staveley-Smith et al.\ 1992).\markcite{stavely1992}
As an evidence for the second scenario, it is argued that the
near-infrared emission in the vast majority of BCDs is attributable to old K
and M giants, which are the major component of dEs (Thuan 1983; Hunter \&
Gallagher 1985).\markcite{thuan1983, hunter1985}

Similarly, there exist two competing hypotheses to explain the evolutionary
connection between dIs and dEs.  The first hypothesis states that dE galaxies
are the faded remnants of previously actively star-forming dI galaxies whose
gas has been lost.  
There exists circumstantial evidence that dEs have in
fact evolved directly from dIs.  Faber \& Lin (1983)\markcite{faber1983} and
Kormendy (1985)\markcite{kormendy1985} have used the similarity in the surface
brightness profiles of dIs and dEs, which are mostly exponential, to argue
that gas-rich dIs are the progenitors of dEs.  The second hypothesis for the
relation between dIs and dEs states that they represent parallel sequences of
dwarf galaxies, fundamentally separated by the intrinsic difference in their
structure.  The observational evidence for this hypothesis is based mostly on
the differences in appearance between the two types of dwarf galaxies; for
instance, dIs have a more diffuse light distribution than dEs, and lack the
bright nucleation which is frequently found in dEs.  In addition to these
differences, there is a dissimilarity in the flattening distribution of dEs
and dIs; the apparent flattening of a galaxy is customarily given either by
the apparent axis ratio $q$ or by the ellipticity $\epsilon \equiv 1 - q$.
Bothun et al.\ (1986) and Impey \& Bothun (1997)\markcite{bothun1986,
  impey1997} presented the results of Ichkawa, Wakamatsu, \& Okamura
(1986)\markcite{ ichikawa1986} and Caldwell (1983)\markcite{caldwell1983} as
evidence for the different flattening distributions between dEs and dIs.
However, the analysis of Ichikawa et al.\ was based on the comparison between
the flattening distributions of dEs and bright (non-dwarf) spiral galaxies;
the situation is similar for Caldwell's analysis.  In addition, contrary to
the claims of Bothun et al.\ and Impey \& Bothun, both Ichikawa et al.\ and
Caldwell showed that the flattening distribution of dEs is similar to that of
bright irregular galaxies.

There have been previous attempts to compare the apparent axis ratio
distributions between LSB dI galaxies and other types of dwarf galaxies.  For
example, Staveley-Smith et al.\ (1992)\markcite{stavely1992}
constructed the axis ratio distribution for
438 Uppsala Galaxy Catalogue (hereafter UGC, Nilson 1973)\markcite{nilson1973}
LSB galaxies, and compared it to that of BCDs whose ellipticities were
measured from Palomar Observatory Sky Survey (POSS) plates by Gorden \&
Gottesman (1981).\markcite{gorden1981} However, previous studies of axis ratio
distributions suffer from large uncertainties for several reasons.  First,
owing to the small dimensions and low surface brightness of dwarf galaxies,
estimating their axis ratio is difficult and leads to large uncertainties.
Second, the UGC sample used by 
Staveley-Smith et al.\ (1992)\markcite{stavely1992} is known to be
inhomogeneous, containing galaxies ranging from true dwarf galaxies to more
luminous very low surface brightness systems (Thuan \& Seitzer 1979; McGaugh,
Schombert, \& Bothun 1995).\markcite{thuan1979, mcgaugh1995} Finally, previous
determinations of LSB dI axis ratios have been based on photographic plates;
for comparison with recent CCD observations of other types of dwarf galaxy, it
is essential to have measurements of the axis ratios of a homogeneous
sample of LSB dIs based on modern CCD observations.

In this paper, we measure the ellipticities of 30 LSB dI galaxies and compare
the ellipticity distribution with that of 80 dEs (Ryden \& Terndrup 1994;
Ryden et al.\ 1998)\markcite{ryden 1994; 1998} and 62 BCDs (Sung et al.\ 
1998).\markcite{sung1998} We find that the ellipticity distribution of LSB dIs
is very similar to that of BCDs, and marginally different from that of dEs.
We then determine, under various assumptions, the distribution of intrinsic
shapes of dI galaxies and compare it to that of other types of dwarfs.
First, we assume that LSB dIs are either all oblate or all prolate, and use
non-parametric analysis to find the best-fitting distribution of intrinsic
shapes. With this assumption, we find that the scarcity of nearly circular LSB
dIs implies, at the 99\% confidence level, that they cannot be a population of
randomly oriented oblate or prolate objects.  Next, we assume that dIs are
triaxial, and use parametric analysis to find permissible distributions of
intrinsic shapes.  We find that if the intrinsic axis ratios, $\beta$ and
$\gamma$, are distributed according to a Gaussian with means $\beta_0$ and
$\gamma_0$ and a common standard deviation of $\sigma$, the best-fitting set
of parameters for LSB dIs is $(\beta_0,\gamma_0,\sigma) = (0.66,0.50,0.15)$,
and the best fit for BCDs is $(\beta_0,\gamma_0,\sigma) = (0.66,0.55,0.16)$,
while the best fit for dEs is $(\beta_0,\gamma_0,\sigma) = (0.78,0.69,0.24)$.
The LSB dIs and BCDs thus have a very similar shape distribution, given this
triaxial hypothesis, while the dEs peak at a somewhat more spherical shape.
Our results are consistent with an evolutionary scenario in which the three
types of dwarf galaxy have a close relation with each other.

\section{Observations}

Our sample consists of 30 LSB dI galaxies which are drawn from the list of UGC
``dwarfs'' and LSB galaxies detected in HI by Schneider et al.\ (1990;
1992).\markcite{schneider1990, schneider1990} In the UGC catalogue, ``dwarfs''
are categorized as ``objects with very low surface brightness and little or no
concentration of light on the red prints'' with Hubble types of Sc-Irr or
later (Nilson 1973).\markcite{nilson1973} Among these galaxies, we select only
galaxies with small 21 cm HI line widths, $\Delta v_{\rm 20} \leq 100\ {\rm km\ 
  s}^{-1}$, small redshifts, $v_0 \leq 1,500\ {\rm km\ s}^{-1}$, and faint
$B$-band luminosities, $M_{B}\gtrsim -16$.  For galaxies not observed by
Schneider et al.\ (1992)\markcite{schneider1992}, HI data were taken from
Huchtmeier \& Richter (1989).\markcite{huchtmeier1989} In addition, we exclude
galaxies with noticeable spiral patterns so that the sample is composed of
pure dwarf irregular galaxies.  In POSS prints, most galaxies in our sample
are found to be of generally low surface brightness, with superimposed
irregular patches of star formation.

The photometric observations of the sample galaxies were carried out during
several observing runs from 1985 to 1993 with different CCD chip and telescope
combinations; by using the KPNO 4\ m with a $320\times512$ RCA1 chip during
1995 May 22--23, the KPNO 2.1\ m with a $512\times512$ T5HA chip during 1990
October 19--22 and 1991 April 17--21, the 2.1\ m with a $1024\times1024$ T1KA
chip during 1993 January 23--24, the KPNO 0.9\ m with a $1024\times1024$ ST1K
chip during 1991 September 13--16, and the 0.9\ m with a $1024\times1024$ T2KA chip
during April 18--20.  The galaxy names and observed bands are listed in Table
1; images of individual galaxies were presented in Figure 2 of Patterson \&
Thuan (1996).\markcite{patterson1996}

All steps of the data reduction and analysis were carried out using a standard
CCD reduction process with IRAF\footnote{IRAF is distributed by National
  Optical Astronomy Observatories, which is operated by the Association of
  Universities for Research in Astronomy, Inc., under cooperative agreement
  with the National Science Foundation.}.  First, a bias offset was subtracted
from each raw frame.  To compensate for pixel-to-pixel variations in the bias
level, we constructed a composite zero frame of 20 individual bias frames
(with the overscan already subtracted), and subtracted it from each frame.
Next, images were divided by the combined flat images constructed from high
signal-to-noise level dome and sky flats for individual nights to remove the
pixel-to-pixel variation in the detector sensitivity.  Then, blank dark sky
exposures were used to remove the interference pattern produced by night sky
emission lines.  Finally, after individual object frames were processed
through the flat fielding correction, separate exposure were aligned and
combined, followed by sky subtraction. Further details of the data reduction
can be found in Patterson \& Thuan (1996).

\section{Axis Ratio Determination}

We determine the apparent axis ratios of individual LSB dIs by fitting
ellipses to the isophotes of obtained images.  For this process, the most
widely used program is the Space Telescope Science Data Analysis System
(STSDAS) routine ``isophote'', which is based on an iterative least-square fit
to a Fourier expansion.  However, since the routine is designed for fitting
the surface brightness distributions of stellar systems with uniform profiles,
such as elliptical galaxies, it often produces an unstable fit when it is used
for spirals and irregular galaxies (Freudling 1992).\markcite{freudling1992}
Therefore, we take a different approach which utilizes a fully two-dimensional
linear fit of the harmonics to the image.  In this approach, the intensity of
a galaxy is parameterized as a series of harmonic terms by
$$
I(a,\psi) = \sum_{n=0}^{k} I_n (a) \cos 
\left\{ n \left[ \psi - \psi_n (a)\right]\right\},
\eqno(3.1)
$$
where $\psi$ is the position angle with respect to the major axis of the
ellipse (Franx, Illingworth, \& Heckman 1989).\markcite{franx1989} The results
of harmonic fits are then used as the initial values for the usual Fourier
series expansion ellipse fitting routine, ``isophote''.  Due to the small
angular size of our images we allowed the program to fit ellipses up to the
radius at which only 60\% of points on the ellipse lay within the image.  The
center and position angle of the isophotes were generally allowed to vary.
During the fitting process we excluded HII regions within the galaxy from each
image along with foreground stars and cosmic rays.  The measured ellipticity
profiles of individual galaxies were presented in Figures 64-71 of Patterson
(1995).\markcite{patterson1995}

The axis ratios of the LSB dI galaxies in our sample, like those of many
stellar systems, vary as a function of semimajor axis.  We find that the
variation is most severe in the inner parts, and is mostly caused by the
existence of irregular structures.  In the outer parts, on the other hand, the
ellipse-fitting process fails as the surface brightness falls far below the
sky brightness.  Therefore, as a representative axis ratio we determine the
intensity-weighted axis ratio averaged over the intermediate region where we
could obtain stable values of axis ratios with successful ellipse-fitting
process.  The intensity-weighted mean axis ratio is computed by
$$
\bar{q} = { \int q(a) dL\over \int dL};\qquad
dL = 2\pi q a \left[ 1 + {1\over 2}
{d {\rm ln}q (a) \over d {\rm ln} a}
\right] \Sigma (a) da,
\eqno(3.2)
$$
where $q = b/a$, $a$ is the semimajor axis of the isophote, $b$ is the
semiminor axis, and $\Sigma (a)$ is the surface brightness of the isophote
with semimajor axis $a$.  In Table 1 we present the finally determined mean
ellipticities, $\bar\epsilon \equiv 1 - \bar{q}$, and their uncertainties.
The errors are estimated by computing the variance of $q (a)$ within the range
of semimajor axes where ellipticities are measured.

Due to the variation of axis ratios within a galaxy, it is important to apply
a consistent method of ellipticity determination for the comparison between
different types of galaxies.  Since the mean ellipticities for the dE sample
of Ryden \& Terndrup (1994) and Ryden et al.\ (1998)\markcite{ryden1994,
ryden1998} and those for BCDs of Sung et al.\ (1998)\markcite{sung1998} were
determined adopting the same method for similarly obtained CCD data, we can
directly compare the axis ratio distribution of LSB dIs with the other types
of dwarf galaxies.  In the upper panel of Figure 1, we present the cumulative
function of $\bar{q}$ for 30 LSB dIs in our sample (solid step function), and
compare it with those of 80 dE and 62 BCD samples.  From this comparison, we
find that the axis ratio distribution of LSB dIs is very similar to that of
BCDs; the Kolmogorov-Smirnov (hereafter KS) probability for comparing these
two samples is $P_{\rm KS} = 0.70$.  Compared to dEs, LSB dIs are slightly
flatter, on average.  For the sample of LSB dIs the mean and standard
deviation of $\bar{q}$ are $0.64 \pm 0.15$, while those for the dE sample are
$0.70 \pm 0.16$.  However, the difference in the axis ratio distributions
between these two samples is marginal with a KS probability of $P_{\rm
  KS}=0.060$.  The results of comparing the apparent axis ratio distributions
between the different types of dwarf galaxies are summarized in Table 2.

\section{Determining Intrinsic Shapes}

For the determination of intrinsic shapes, we apply two different methods:
non-parametric and parametric.  The non-parametric method assumes that the
galaxies in a sample are either all oblate or all prolate, with intrinsic axis
ratio $\gamma$, and are randomly oriented relative to us.  With these
assumptions, the distribution $f(\gamma)$ of intrinsic shapes can be determined
from the distribution $f(q)$ of apparent shapes by performing a unique
mathematical inversion.  The parametric method, by contrast, assumes that the
galaxies are triaxial, with axis lengths in the ratio $1 : \beta : \gamma$,
where $1 \geq \beta \geq \gamma$. In this case, there
is no longer a unique inversion from the observed distribution
$f(q)$ to the intrinsic distribution $f(\beta,\gamma)$. However, using a
parametric model for $f (\beta,\gamma)$, we can find the model distribution of
intrinsic axis ratios which best fits the observed distribution of apparent
axis ratios.  Compared to the first method, the parametric method has the
disadvantage that one has to assume a functional form (e.g., Gaussian) for the
model axis ratio distribution, which is actually poorly known.  Nevertheless,
parametric fits are useful because they show us how statistics such as the KS
and $\chi^2$ scores vary as the parameters are changed.

\subsection{Non-parametric Method}

If all LSB dI galaxies were randomly oriented oblate spheroids and if we knew
exactly the distribution of apparent shapes $f(q)$, then we could perform a
mathematical inversion on $f(q)$ to find the distribution $f_o (\gamma)$ for
the intrinsic axis ratios of the oblate spheroids. Unfortunately, we don't
know $f(q)$; we only have a sample of finite size drawn from the distribution.
Thus, we can only accept or reject, at a known confidence level, the null
hypothesis that the galaxies in the sample are randomly oriented oblate
spheroids. In addition, if the null hypothesis is not rejected, we can present
an estimate of the distribution function $f_o (\gamma)$. In a similar way, we
can accept or reject the hypothesis that the galaxies in the sample are
randomly oriented prolate spheroids.

To test the oblate and prolate hypotheses, we start by making a non-parametric
kernel estimate of the distribution $f(q)$ of the apparent axis ratios.
Details of how non-parametric kernel estimators are used in this context are
given by Tremblay \& Merritt (1995)\markcite{tremblay1995} and Ryden
(1996).\markcite{ryden1996} We then numerically invert our estimate for $f(q)$
to find estimates for $f_o (\gamma)$ and $f_p (\gamma)$, the distributions of
intrinsic axis ratios given the oblate and prolate hypotheses, respectively.
Confidence intervals are placed on the estimates for $f$, $f_o$, and $f_p$ by
performing repeated bootstrap resampling of the original data set and creating
new estimates from each bootstrap resampling. The spread in the bootstrap
estimates of $f$ at a given value of $q$, and in the bootstrap estimates of
$f_o$ and $f_p$ at a given value of $\gamma$, provides confidence intervals
for the non-parametric estimates of these functions. Once the estimates for
$f_o$ and $f_p$ are determined, we can reject the oblate or prolate hypothesis
at a given confidence level if the upper confidence level drops below zero for
any value of the intrinsic axis ratio $\gamma$.  (A hypothesis that calls for
a negative number of galaxies at a given axis ratio is unphysical, and should
be firmly rejected.)

In the upper panel of Figure 2, we present the non-parametric kernel estimate
of the distribution of the apparent axis ratios $\bar{q}$ for our sample of 30
LSB dIs. In the middle panel, we show the distribution of intrinsic axis
ratios assuming the LSB dIs are oblate; in the lower panel, we show the
distribution of intrinsic axis ratios assuming they are prolate.  In each
panel, the solid line is the best estimate, the dashed lines show the 80\%
confidence band, and the dotted lines show the 98\% confidence band. (That is,
at a given value of $q$, 1\% of the bootstrap estimates fall above the upper
boundary of the 98\% confidence band, and 1\% fall below the lower boundary of
the 98\% confidence band.) A Gaussian kernel was used to ensure a smooth,
differentiable estimate of $f$, with a width $h = 0.069$. Because we imposed a
more-or-less arbitrary reflective boundary condition at $q = 1$, we don't
believe our estimates for $f$, $f_o$, and $f_p$, within a distance $\sim h$ of
the right-hand edge of Figure 2.

Note that there is a decided lack of nearly circular LSB dIs in our sample;
the roundest galaxy we observed has $q = 0.873$. This scarcity of nearly
circular galaxies is the characteristic sign that the galaxies cannot be a
population of oblate spheroids. Looking at the middle panel, we see that the
oblate hypothesis can be ruled out at the 99\% (one-sided) confidence level.
The 98\% confidence band for $f_o$ drops below zero for axis ratios $\gamma >
0.85$.  Indeed, so pronounced is the lack of nearly circular galaxies, even
the prolate hypothesis can be ruled out at the 99\% (one-sided) confidence
level. The 98\% confidence band for $f_p$ drops below zero for $\gamma >
0.90$.  Thus, even with our relatively small sample of galaxies, we can reject
at a high confidence level the hypothesis that the LSB dI galaxies are a
population of randomly oriented spheroids, either oblate or prolate. This
leads us to consider, in the next section, possible distributions of triaxial
shapes for the LSB dI galaxies in our sample.

\subsection{Parametric Method}

To determine the intrinsic axis ratio distribution of LSB dI galaxies, we
model the distribution of intrinsic axis ratios $(\beta, \gamma)$ as a
Gaussian distribution with means $\beta_0$ and $\gamma_0$ and a common width
$\sigma$; i.e.,
$$
f(\beta,\gamma )  \propto
\exp \left[ -
{(\beta-\beta_0)^2 + (\gamma-\gamma_0)^2
\over
2\sigma^2
}
\right].
\eqno(4.2.1)
$$
We then produce a large number ($\sim 10^5$) of test galaxies with their
intrinsic axis ratios distributed according to equation (4.2.1), and with
random orientations with respect to us. 
Once test galaxies are produced, we then compute their projected axis
ratio $q$. When a triaxial ellipsoid is projected with
the viewing angles of $\theta$ and $\phi$, it appears as an ellipse with an
apparent axis ratio of
$$
q(\beta,\gamma,\theta,\phi) =
\left[ {
A + C - \sqrt{(A-C)^2+B^2}
\over
A + C + \sqrt{(A-C)^2+B^2}
} \right]^{1/2},
\eqno(4.2.2)
$$
where
$$
\cases{
A = (\cos^2 \phi + \beta^2\sin^2\phi)\cos^2\theta + \gamma^2\sin^2\theta,\cr
B = \cos\theta\sin 2\phi (1-\beta^2),\cr
C = \sin^2\phi + \beta^2\cos^2\phi
}
\eqno(4.2.3)
$$
(Binney 1985).\markcite{binney1985} We then statistically compare the
projected axis ratio distribution of test galaxies with the observed one using
a KS test (for the cumulative distribution) and $\chi^2$ test (for the binned
distribution).

In Figure 3a and 3b, we present the isoprobability contours on 6 slices
through the $(\beta_0,\gamma_0,\sigma)$ parameter space, as measured by KS and
$\chi^2$ tests, respectively, for our LSB dI sample.  When measured by a KS
test, the best-fitting distribution has parameters of
$(\beta_0,\gamma_0,\sigma) = (0.66, 0.50, 0.15)$ with the KS probability
$P_{\rm KS} = 0.98$, implying that the intrinsic shape of LSB dIs can be well
fitted by a population of triaxial ellipsoids.  We obtain consistent results
when measured by $\chi^2$ tests; the best-fitting distribution has parameters
of $(\beta_0,\gamma_0,\sigma) = (0.80, 0.42, 0.20)$ with the $\chi^2$
probability $P_{\chi^2} = 0.91$.

For the comparison with other types of dwarf galaxies, we list the parameters
of the best-fitting distributions for BCDs and dEs in Table 2.  The isoprobability
contours from which these parameters are drawn, were presented in Fig 6a, 6b,
and 7 of Sung et al.\ (1998).  In addition, in the lower panel of Figure 1 we
present the computed cumulative distribution (smooth curve) for the
best-fitting triaxial model, as measured by the KS test.  The best-fitting
model is overlaid on the measured cumulative distribution (step function) for
each type of dwarf galaxy.

\section{Summary}

We measure the ellipticities for a sample of 30 LSB dIs and compare the
distribution of ellipticities with those for the samples of 62 BCDs and 80
dEs.  From this comparison, we find that the axis ratio distribution of LSB
dIs is very similar to that of BCDs.  Compared to dEs, LSB dIs are slightly
flatter, but the difference is marginal.  We also determine the intrinsic
shape of LSB dIs from the distribution of apparent axis ratios.  From the
non-parametric analysis, we find the hypothesis that our sample LSB dIs are
randomly oriented oblate or prolate objects is rejected with strong confidence
level.  On the other hand, the shape of LBS dI galaxies are well described by
triaxial spheroids if their axis ratios, $\beta$ and $\gamma$, have a Gaussian
distribution.  From the parametric analysis, we determine the best-fitting
parameters are $(\beta_0,\gamma_0,\sigma) = (0.66, 0.50, 0.15)$.  These
results directly contradict the long-standing belief that LSB dIs have very
flattened disky shapes, quite different from the spheroidal shapes of dEs
and BCDs.  Therefore, our results are consistent with the scenario that the
three major types of dwarf galaxies have very close evolutionary connections.

\acknowledgements
E.-C.\ S.\ has been supported by Basic Research Fund of Korea
Astronomy Observatory.
B.\ S.\ R.\ was supported by grant NSF AST-93-577396.

\clearpage

\begin{center}
\bigskip
\bigskip
\centerline{\small {TABLE 1}}
\smallskip
\centerline{\small {\sc The Intensity Weighted Mean Ellipticities}}
\smallskip
\begin{tabular}{clcl|clcl}
\hline
\hline
\multicolumn{1}{c}{No.} &
\multicolumn{1}{c}{galaxy} &
\multicolumn{1}{c}{obs.} &
\multicolumn{1}{c}{ellipticity} &
\multicolumn{1}{|c}{No.} &
\multicolumn{1}{c}{galaxy} &
\multicolumn{1}{c}{obs.} &
\multicolumn{1}{c}{ellipticity} \\
\multicolumn{1}{c}{} &
\multicolumn{1}{c}{name} &
\multicolumn{1}{c}{band} &
\multicolumn{1}{c}{$\bar{\epsilon} \pm \Delta\epsilon$} &
\multicolumn{1}{|c}{} &
\multicolumn{1}{c}{name} &
\multicolumn{1}{c}{band} &
\multicolumn{1}{c}{$\bar{\epsilon} \pm \Delta\epsilon$} \\
\hline
1    & KARA 10     & $I$   & 0.429 $\pm$ 0.0809    & 16    & UGC 04173    & $I$   & 0.601 $\pm$ 0.0082 \\    
2    & M81dwA      & $I$   & 0.159 $\pm$ 0.3767    & 17    & UGC 05423    & $I$   & 0.375 $\pm$ 0.0843 \\
3    & UGC 00031   & $I$   & 0.271 $\pm$ 0.1166    & 18    & UGC 07548    & $I$   & 0.471 $\pm$ 0.1832 \\
4    & UGC 00063   & $I$   & 0.421 $\pm$ 0.0454    & 19    & UGC 07596    & $I$   & 0.568 $\pm$ 0.0606 \\
5    & UGC 00300   & $I$   & 0.182 $\pm$ 0.1562    & 20    & UGC 07636    & $I$   & 0.304 $\pm$ 0.0780 \\
6    & UGC 00772   & $B$   & 0.426 $\pm$ 0.0677    & 21    & UGC 07684    & $I$   & 0.368 $\pm$ 0.0267 \\
7    & UGC 01171   & $I$   & 0.383 $\pm$ 0.0340    & 22    & UGC 08091    & $I$   & 0.333 $\pm$ 0.3350 \\
8    & UGC 01981   & $I$   & 0.163 $\pm$ 0.0356    & 23    & UGC 08201    & $B$   & 0.243 $\pm$ 0.3216 \\
9    & UGC 02017   & $I$   & 0.434 $\pm$ 0.1183    & 24    & UGC 08683    & $I$   & 0.415 $\pm$ 0.1102 \\
10   & UGC 02034   & $I$   & 0.127 $\pm$ 0.2909    & 25    & UGC 08760    & $I$   & 0.758 $\pm$ 0.1550 \\
11   & UGC 02053   & $I$   & 0.629 $\pm$ 0.2038    & 26    & UGC 08833    & $I$   & 0.251 $\pm$ 0.0767 \\
12   & UGC 02162   & $I$   & 0.245 $\pm$ 0.1463    & 27    & UGC 09128    & $I$   & 0.481 $\pm$ 0.0993 \\
13   & UGC 03212   & $I$   & 0.587 $\pm$ 0.1721    & 28    & UGC 10031    & $I$   & 0.283 $\pm$ 0.2665 \\
14   & UGC 03817   & $I$   & 0.297 $\pm$ 0.3201    & 29    & UGC 10669    & $I$   & 0.227 $\pm$ 0.1074 \\
15   & UGC 03966   & $I$   & 0.250 $\pm$ 0.1087    & 30    & UGC 12894    & $I$   & 0.277 $\pm$ 0.0633 \\
\hline
\end{tabular}
\end{center}
\smallskip
\noindent
{\footnotesize \qquad NOTE.--- The intensity weighted mean ellipticities of 30 
  LSB dI galaxies and their uncertainties.  The errors are estimated by
  computing the variance of $\epsilon(a)$ within the range of semimajor axis
  $a$ where the ellipticities are measured.  Also marked are the band of
  images for which the ellipticities are measured.  } \clearpage

\begin{center}
\bigskip
\bigskip
\centerline{\small {TABLE 2}}
\smallskip
\centerline{\small {\sc Comparison of Ellipticity Distributions}}
\smallskip
\begin{tabular}{ccccc}
\hline
\hline
\multicolumn{4}{c}{galaxy type} &
\multicolumn{1}{c}{$P_{\rm KS}$} \\
\multicolumn{1}{c}{type 1} &
\multicolumn{1}{c}{$\langle \bar{q}\rangle$} &
\multicolumn{1}{c}{} &
\multicolumn{1}{c}{type 2} &
\multicolumn{1}{c}{} \\
\hline
LSB dIs  & $0.64\pm 0.15$ & vs. & BCDs       & 0.701 \\
BCDs     & $0.67\pm 0.15$ & vs. & dEs        & 0.057 \\
dEs      & $0.72\pm 0.16$ & vs. & LSB dIs    & 0.060 \\
\hline
\end{tabular}
\end{center}
\smallskip
\noindent
{\footnotesize \qquad NOTE.--- Comparisons of observed axis ratio
  distributions between the different types of dwarf galaxies, as measured by
  KS test.  The mean axis ratios $\langle \bar{q}\rangle$ are based on the
  samples of 30 LSB dIs from this paper, 62 BCDs from Sung et al.\ (1998), and
  80 dEs from Ryden \& Terndrup (1994) and Ryden et al.\ (1998).  }

\bigskip
\bigskip
\bigskip
\bigskip

\begin{center}
\bigskip
\centerline{\small {TABLE 3}}
\smallskip
\centerline{\small {\sc Best-Fitting Parameters}}
\smallskip
\begin{tabular}{lccccl}
\hline
\hline
\multicolumn{1}{c}{galaxy} &
\multicolumn{1}{c}{statistical} &
\multicolumn{3}{c}{best-fitting parameters} &
\multicolumn{1}{c}{best-fitting} \\
\multicolumn{1}{c}{type} &
\multicolumn{1}{c}{test} &
\multicolumn{1}{c}{$\beta_{0}$} &
\multicolumn{1}{c}{$\gamma_{0}$} &
\multicolumn{1}{c}{$\sigma$} &
\multicolumn{1}{c}{statistics} \\
\hline
LSB dIs & KS       & 0.66  & 0.50  & 0.15  & $P_{\rm KS}=0.98$ \\
\bigskip
        & $\chi^2$ & 0.80  & 0.42  & 0.20  & $P_{\chi^2}=0.91$  \\
BCDs    & KS       & 0.66  & 0.55  & 0.16  & $P_{\rm KS}=0.99$  \\
\bigskip
        & $\chi^2$ & 0.77  & 0.51  & 0.16  & $P_{\chi^2}=0.96$  \\
dEs     & KS       & 0.78  & 0.69  & 0.24  & $P_{\rm KS}=0.99$  \\
        & $\chi^2$ & 0.87  & 0.64  & 0.24  & $P_{\chi^2}=0.94$  \\
\hline
\end{tabular}
\end{center}
\smallskip
\noindent
{\footnotesize \qquad NOTE.--- Best-fitting parameters and statistics for
  intrinsic axis ratio distributions for the three different types of dwarf
  galaxies, as measured by KS and $\chi^2$ tests.  We assume that galaxies are
  triaxial ellipsoids with axis ratio $1 \leq \beta \leq \gamma$, and the
  distribution of intrinsic axis ratios follows a Gaussian distribution with
  means $\beta_0$ and $\gamma_0$ and a common width $\sigma$.  }

\bigskip
\postscript{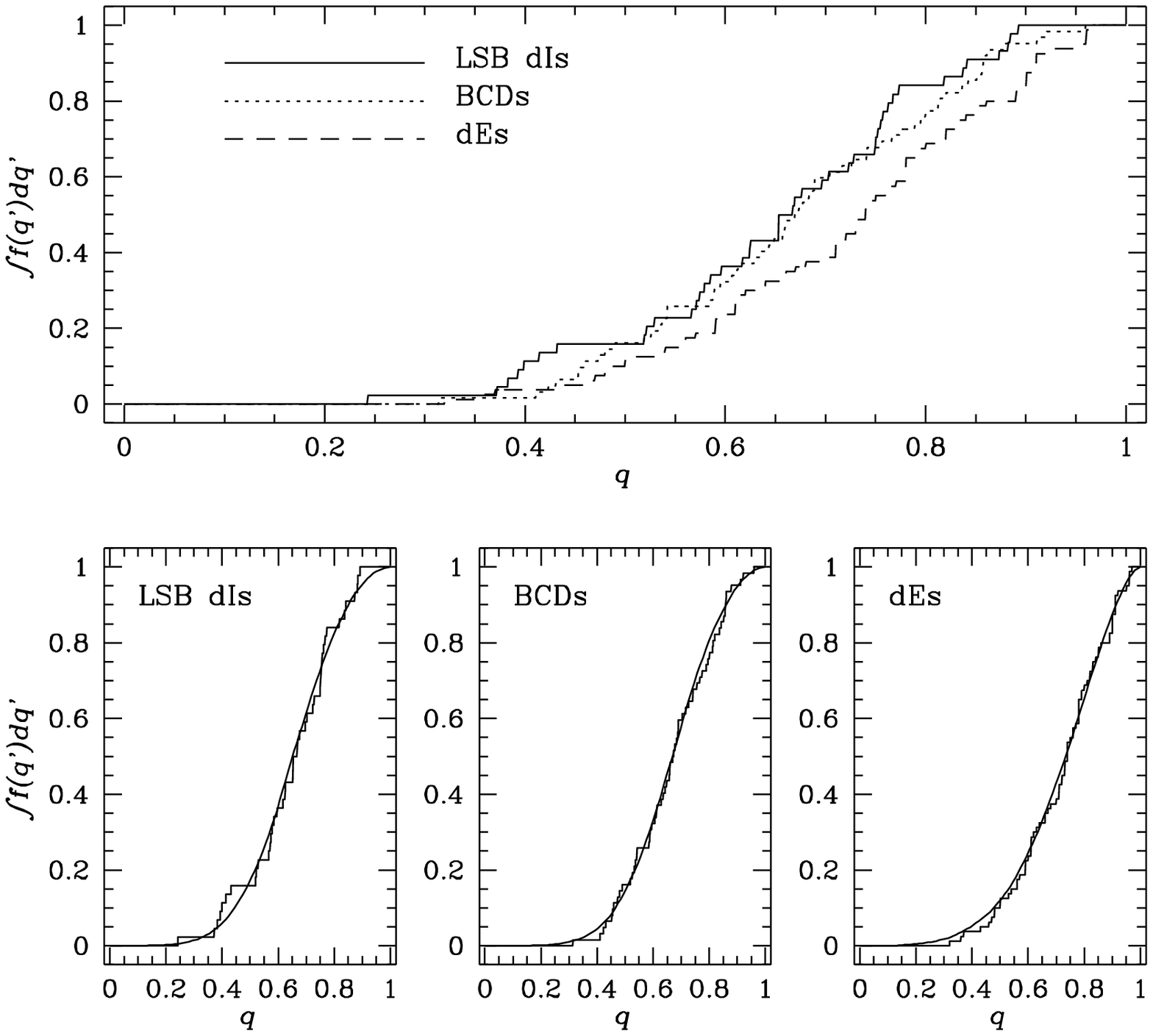}{1.1}
\noindent
{\footnotesize {\bf Figure 1:}\ Upper panel: The cumulative distributions of
  apparent axis ratios for 30 LSB dIs (solid line), 62 BCDs (short-dashed
  line), and 80 dEs (long-dashed line).  Lower panels: The cumulative
  distributions of apparent axis ratio distributions are superimposed on the
  predicted distributions of apparent shapes from the best-fitting triaxial
  models, as measured by  KS tests.  } \clearpage

\bigskip
\postscript{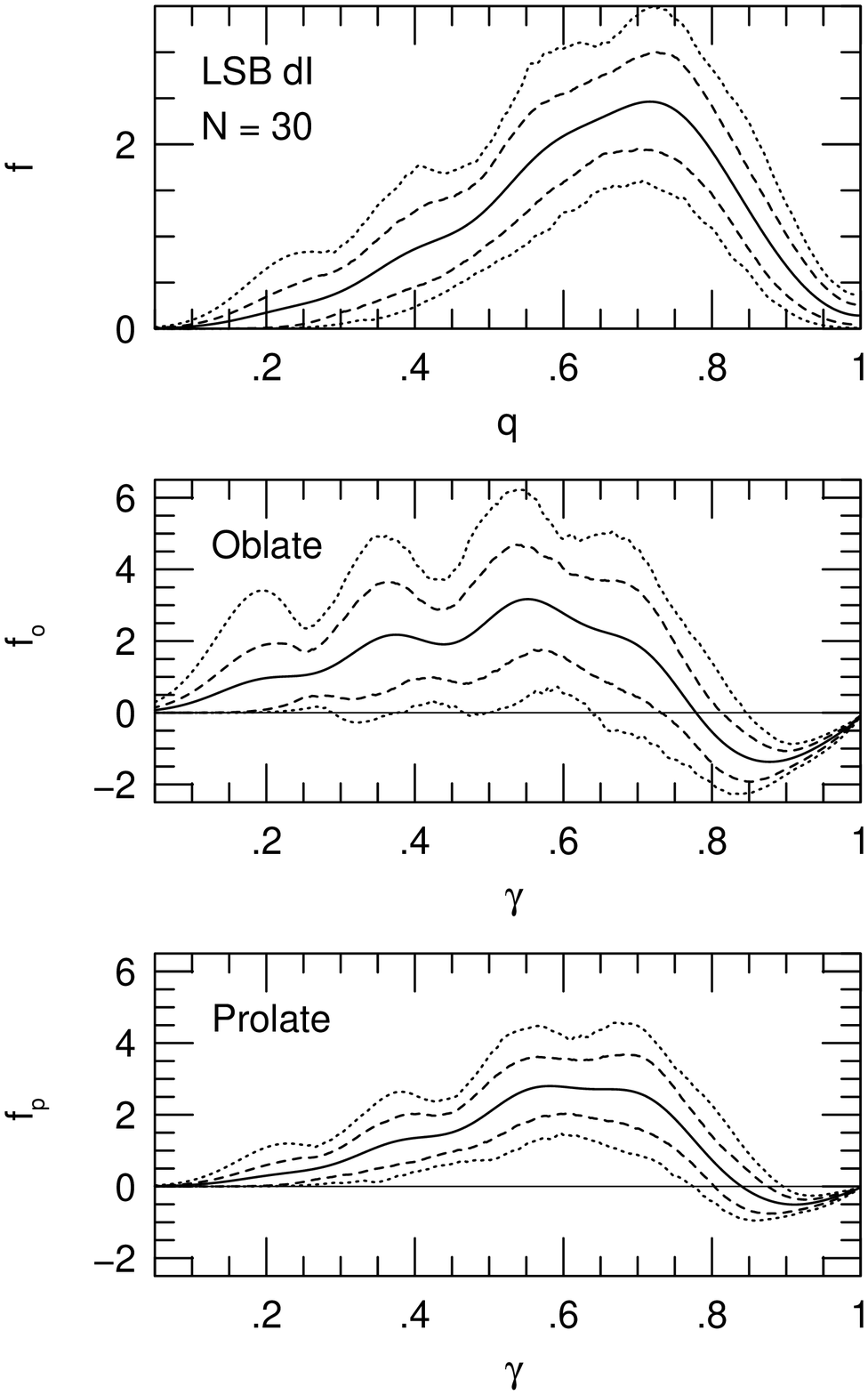}{0.8}
\noindent
{\footnotesize {\bf Figure 2:}\ The non-parametric kernel estimate of the
  distribution of 30 LSB dI sample galaxies (top panel).  Also shown are the
  distributions of intrinsic axis ratios, which are produced under the
  assumption that LSB dIs are all oblate (middle panel) and all prolate
  (bottom panel).  The solid line in each panel is the best estimate, the
  dashed lines are the 80\% confidence band, and dotted lines are the 98\%
  confidence band.} \clearpage

\bigskip
\postscript{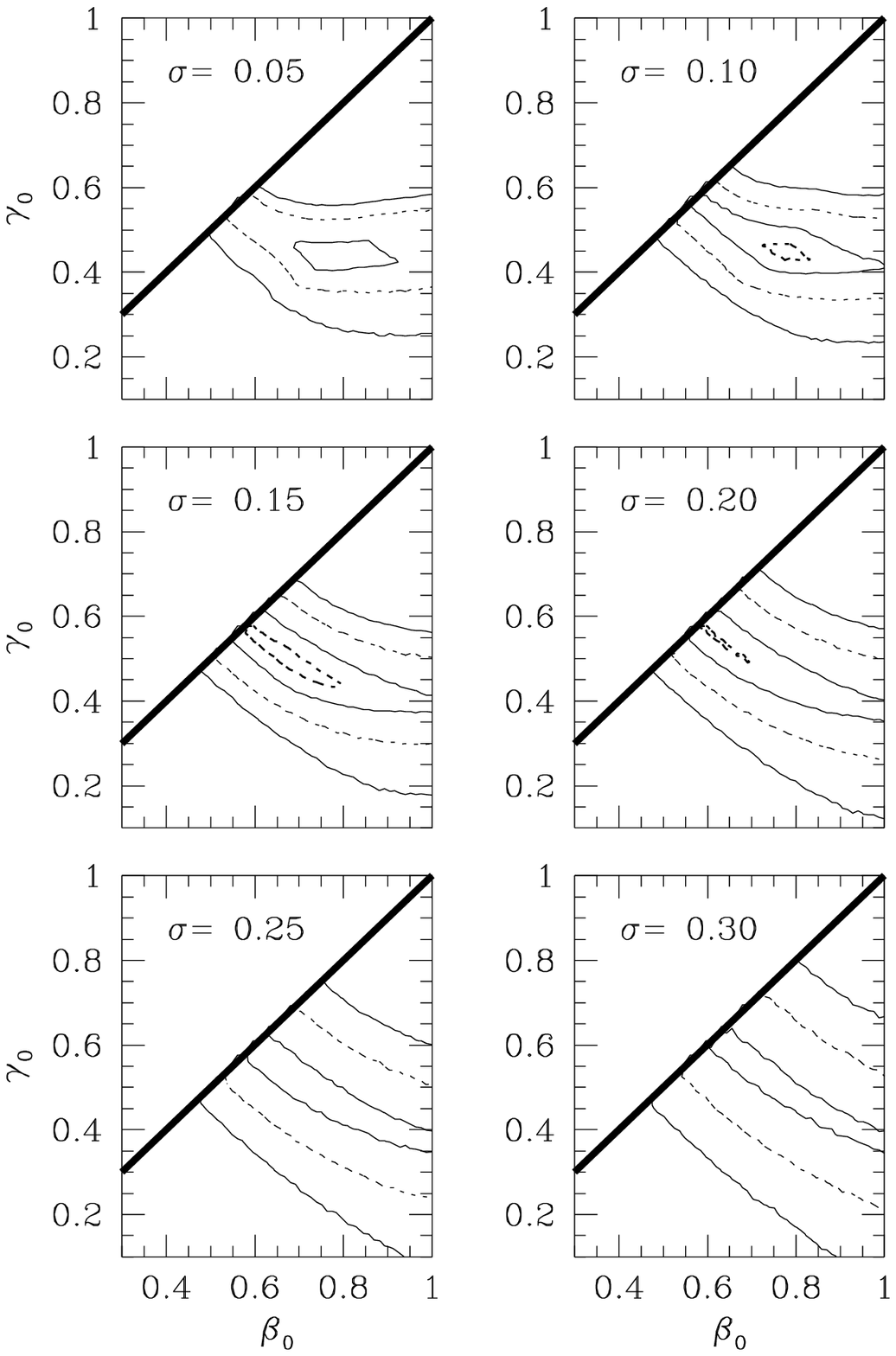}{1.1}
\noindent
{\footnotesize {\bf Figure 3a:}\ The isoprobability contours, as measured by
  KS tests for 30 LSB dIs, on 6 slices through $(\beta_0, \gamma_0, \sigma)$
  parameter space.  Contours are drawn at the levels $P_{\rm KS} = 0.01$,
  $0.1$, $0.5$, and $0.9$, starting from the outside.  } \clearpage

\bigskip
\postscript{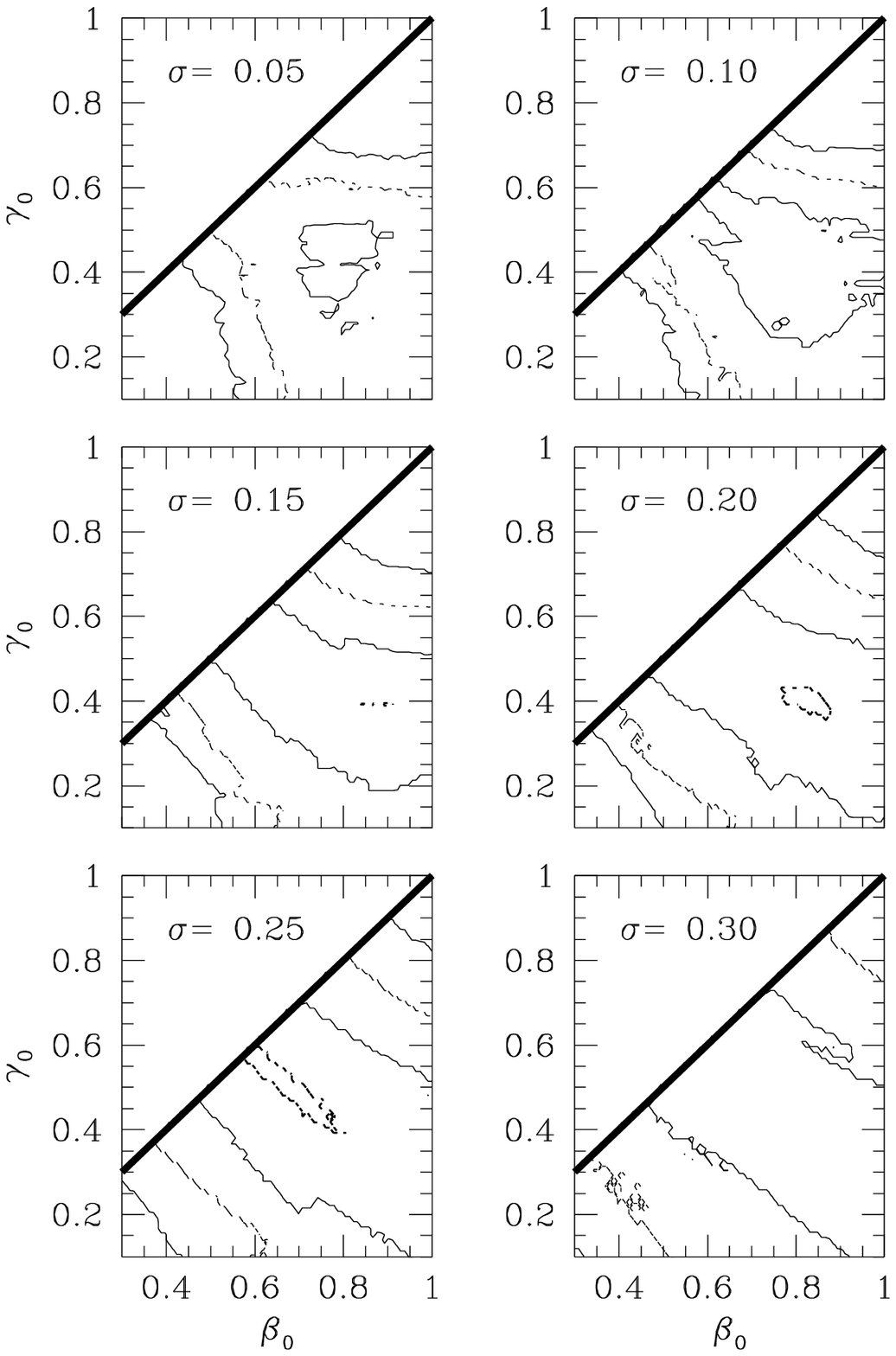}{1.1}
\noindent
{\footnotesize {\bf Figure 3b:}\ The isoprobability contours, as measured by
  $\chi^2$ tests for 30 LSB dIs, on 6 slices through $(\beta_0, \gamma_0,
  \sigma)$ parameter space.  Contours are drawn at the levels $P_{\chi^2} =
  0.01$, $0.1$, $0.5$, and $0.9$, starting from the outside.  } \clearpage

\end{document}